\documentclass{moriond}

\def\be{\begin{equation}}
\def\ee{\end{equation}}
\def\bea{\begin{eqnarray}}
\def\eea{\end{eqnarray}}

\newcommand{\Photo}{\includegraphics[height=35mm]{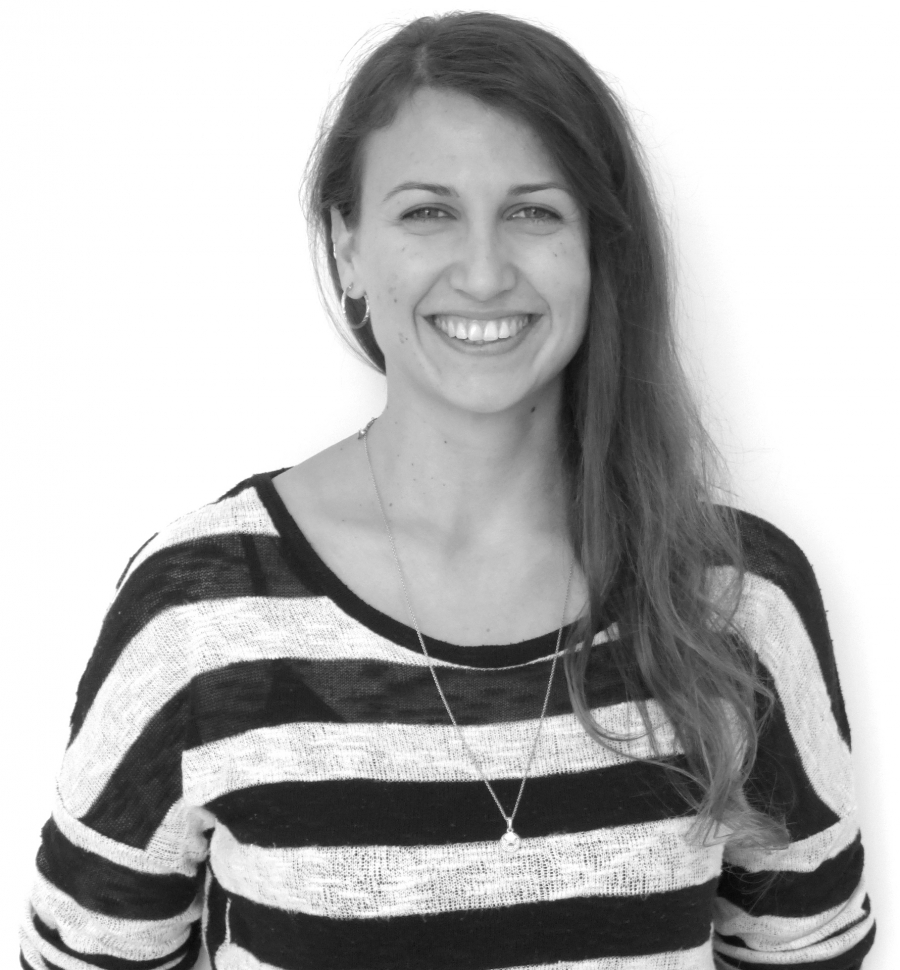}}

\begin{document}
\vspace*{4cm}
\title{LATEST RESULTS FROM THE DAMPE SPACE MISSION}

\author{ FRANCESCA ALEMANNO ON BEHALF OF THE DAMPE COLLABORATION }

\address{Gran Sasso Science Institute (GSSI),\\ Via Iacobucci 2, I-67100 L’Aquila, Italy\\
 and Istituto Nazionale di Fisica Nucleare (INFN), \\ Laboratori Nazionali del Gran Sasso, I-67100 Assergi, L’Aquila, Italy.}

\maketitle\abstracts{
The DArk Matter Particle Explorer (DAMPE) is a space-based particle detector launched on December 17th, 2015 from the Jiuquan Satellite Launch Center (China). The main goals of the DAMPE mission are the study of galactic cosmic-rays (CR), the electron-positron energy spectrum, gamma-ray astronomy and indirect dark matter search. Among its sub-detectors, the deep calorimeter makes DAMPE able to measure electrons and gamma-ray spectra up to 10 TeV, and CR nuclei spectra up to hundreds of TeV, with unprecedented energy resolution. This high-energy region is important in order to search for electron-positron sources, for dark matter signatures in space, and to clarify CR acceleration and propagation mechanisms inside our galaxy. A general overview of the DAMPE experiment will be presented in this work, along with its main results and ongoing activities.}

\section{Introduction}\label{intro}
\begin{figure}[htpb]
\centering
\centerline{\includegraphics[scale=0.23]{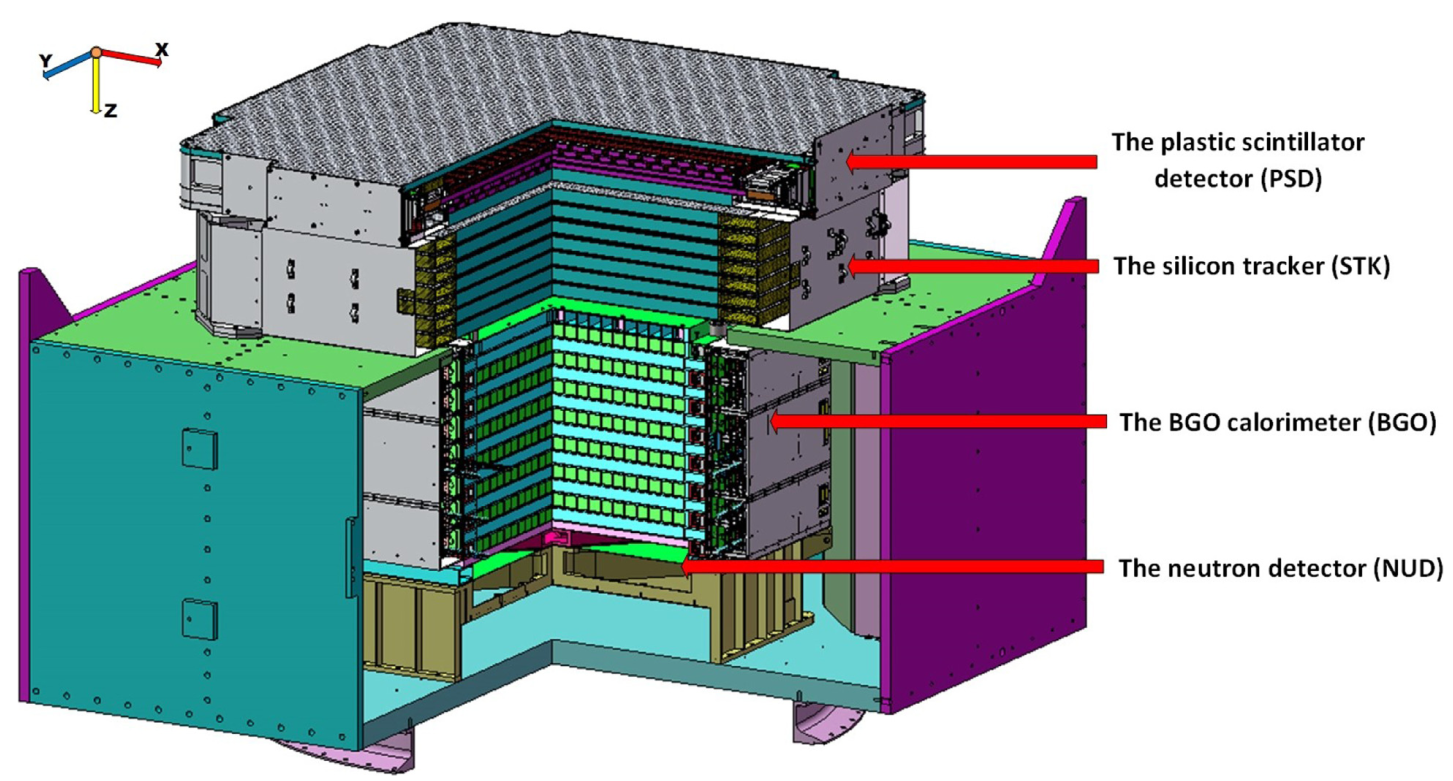}}%
\caption{The DAMPE instrument with its sub-detectors.}
\label{fig:DETECTOR}
\end{figure}

Over the last years, particle measurements in space led to very precise, unexpected results. Deviations from the standard single power-law model have been found in recent observations regarding CR electron and nuclei spectra. High energy resolution of space-borne instruments allows the search of possible indirect dark matter (DM) signatures such as sharp line features. Furthermore, good angular resolution gives the possibility of identifying gamma-ray sources in the sky. In this scenario, the DAMPE experiment was designed in order to contribute to the aforementioned scientific topics. The DAMPE instrument is composed by 4 sub-detectors (see figure \ref{fig:DETECTOR}). Starting from the top, the plastic scintillator detector (PSD) is used for charge measurement and electron-gamma separation. Below, a silicon tungsten tracker (STK) is instrumented, composed of alternated silicon-tungsten layers for precise particle tracking and to induce gamma-ray pair production. After the STK, a BGO calorimeter (BGO) of 32 radiation lengths and 1.6 nuclear interaction lengths is placed, for the purpose of performing precise energy measurement and hadronic/electromagnetic shower discrimination; at the bottom, the neutron detector (NUD) provides an additional level of hadron rejection \cite{9}. Due to its deep calorimeter, DAMPE can measure gamma-rays and CR electrons+positrons up to tens of TeV, as well as CR nuclei spectra up to a few hundreds of TeV. By exploring such high energy regions, with optimal resolution, DAMPE can provide precise CR spectral measurements, thus contributing to the comprehension of their acceleration and propagation mechanisms in our galaxy, while searching for a possible indirect dark matter signature.

\section{ELECTRON-POSITRON SPECTRUM}
\begin{figure}[htpb]
\centering
\centerline{\includegraphics[scale=0.23]{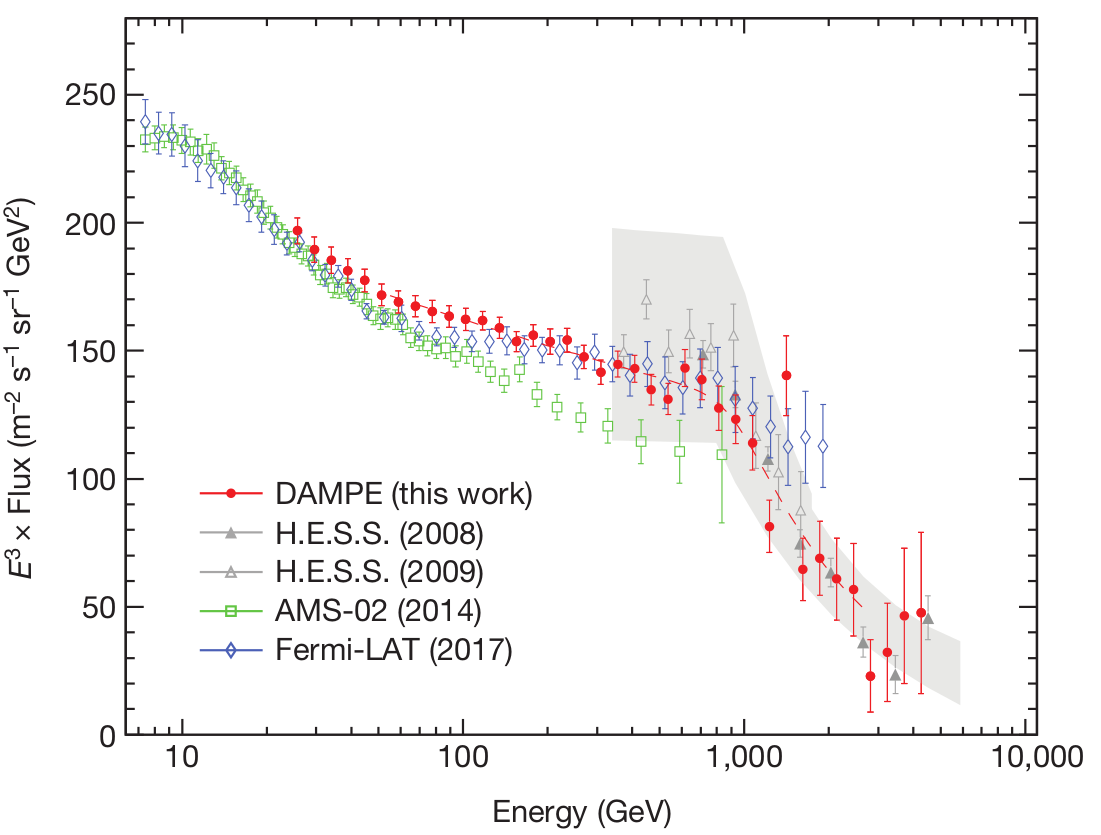}}%
\caption{The electron-positron energy spectrum measured by DAMPE (red circles) compared with other experimental results \protect\cite{14,15,22,23}. Error bars in the DAMPE spectrum represent both systematic and statistical uncertainties added in quadrature, while the red dashed line shows the fit performed with a smoothly broken power-law function.}
\label{fig:ELECTRONS}
\end{figure}

In 2017, the DAMPE collaboration published the measurement of the electron-positron energy spectrum with unprecedented high energy resolution and very low background \cite{12}. Figure \ref{fig:ELECTRONS} shows the electron-positron spectrum from 25 GeV to 4.6 TeV. The red dashed line represents the fit with a smoothly broken power-law model, favored over a single power-law model. The fit shows, for the first time, a distinct spectral break at an energy of $\sim$0.9 TeV, with small uncertainties, confirming what has already been suggested from previous indirect experiments \cite{14,15}. The analysis of the all-electron spectrum is continuing with more years of accumulated statistics.

\section{GAMMA-RAY SOURCES}
\begin{figure}[htpb]
\centering
\centerline{\includegraphics[scale=0.26]{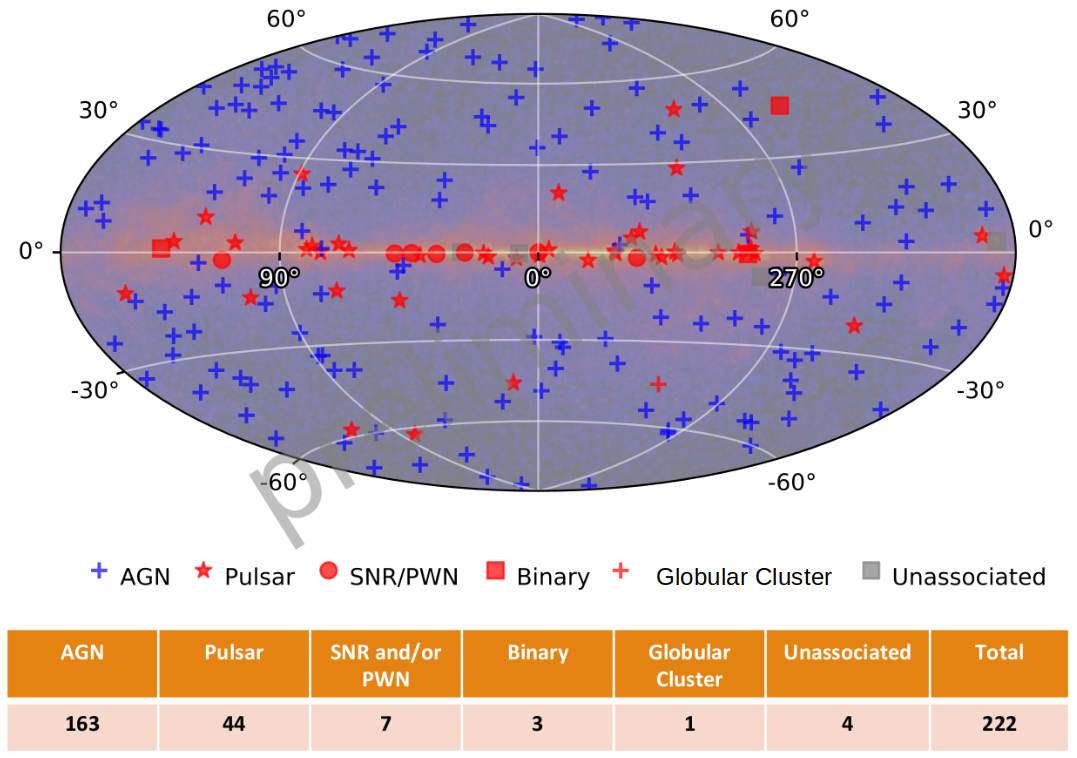}}%
\caption{Gamma-ray sources observed by DAMPE, after association with the 4FGL.}
\label{fig:GSources}
\end{figure}

The gamma-ray analysis has been carried out using 5 years of observed data, resulting in more than 220000 photons detected above 2 GeV \cite{31}. From this preliminary analysis, 222 gamma-ray sources are identified, the majority of which favor a power-law spectral interpretation. In addition, these sources have been associated to the Fermi catalog (4FGL \cite{32}), leading to a source association with Active Galactic Nuclei (AGNs) and pulsars (figure \ref{fig:GSources}).\\

\section{GAMMA-RAY LINES AND DARK MATTER SEARCH}
\begin{figure}[htpb]
\centering
\centerline{\includegraphics[scale=0.25]{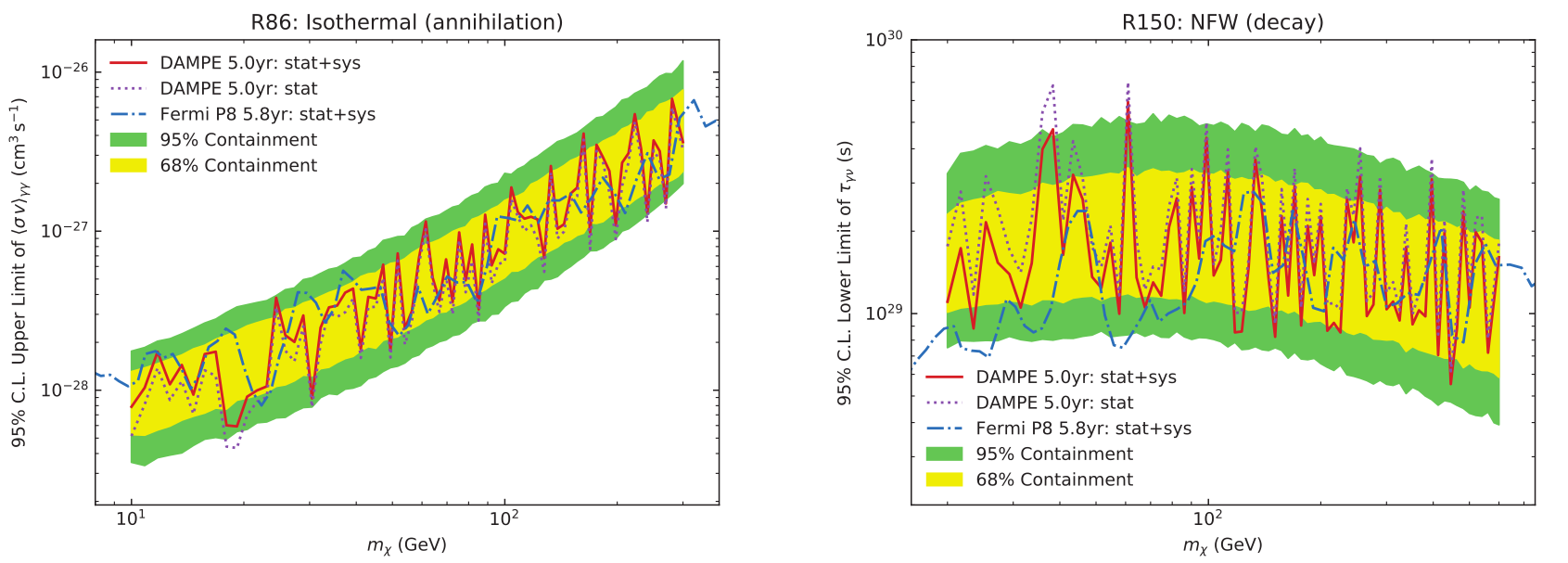}}%
\caption{95\% confidence level upper limits for annihilating dark matter (left panel) and lower limits for lifetime of decaying dark matter (right panel) obtained with 5-years of DAMPE data, with and without systematic uncertainties (red solid and purple dotted lines) compared with the 5.8-year Fermi-LAT results (blue dot-dashed line). Yellow and green bands show the 68\% and 95\% expected containment.}
\label{fig:DM}
\end{figure}

Many theories have been proposed to explain the presence of dark matter in our universe. Among them, the WIMP hypothesis, postulates that dark matter could be made of weakly interactive massive particles (WIMP). If two WIMPs can annihilate into a photon and another particle, it should be, in principle, possible to see a monochromatic structure in the gamma-ray spectrum at energy $E_{\gamma}=m_{\chi}\left(1-m_{X}^{2} / 4 m_{\chi}^{2}\right)$, where $m_{\chi}$ and $m_{X}$ are the mass of the WIMP and of the standard model particle respectively.
Among the main goals of the DAMPE mission there is the search for dark matter particles in space, which could eventually interact or decay, yielding to a detectable signal. A search for gamma-ray lines has been performed using 5 years of data from the DAMPE satellite, focusing on the gamma-ray spectrum from 10 GeV to 300 GeV. Although no DM candidates were found, the analysis provided 95\% confidence level constraints on the DM decay lifetime and annihilation cross section \cite{11}. In figure \ref{fig:DM}, the 5-year DAMPE results are compared with the 5.8-year Fermi-LAT results \cite{10}. The excellent energy resolution of DAMPE improved the limits for decaying DM with mass $\leq$ 100 GeV, with respect to the Fermi-LAT ones.

\section{COSMIC RAYS}
\begin{figure}[htpb]
  \centering
  \centerline{\includegraphics[scale=0.18]{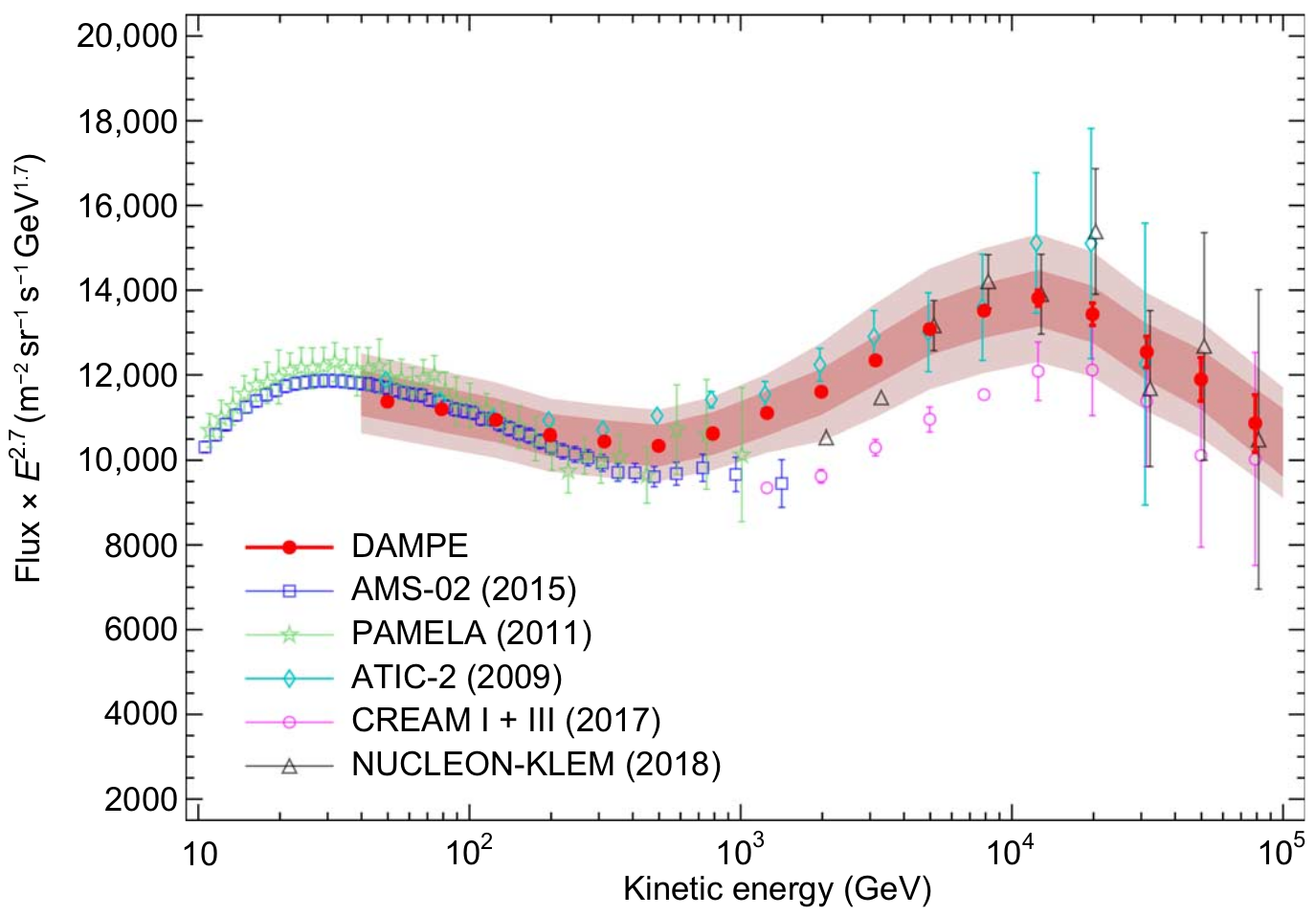}}%
  \caption{DAMPE proton spectrum (red circles), compared with previous direct measurements \protect\cite{1,2,3,5,7,16}. Statistical uncertainties are represented by error bars, while the inner and outer shaded bands indicate systematic uncertainties due to the analysis procedure and total ones (including the hadronic model) respectively.}
\label{fig:Pr}
\end{figure}

\begin{figure}[htpb]
  \centering
  \centerline{\includegraphics[scale=0.27]{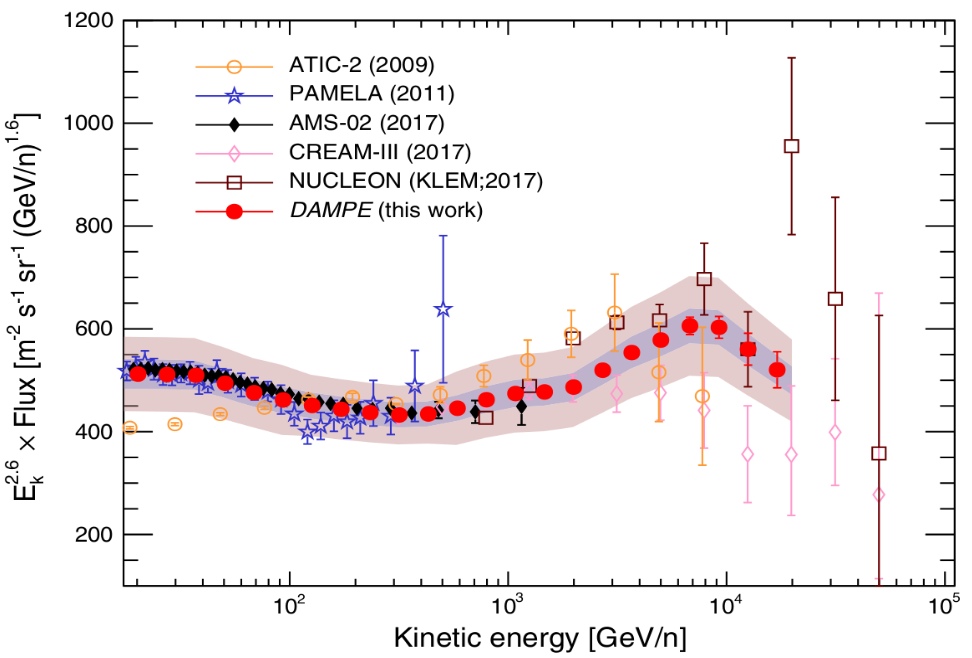}}%
  \caption{DAMPE helium spectrum (red circles), compared with previous direct measurements \protect\cite{1,2,3,5,7,16}. Statistical uncertainties are represented by error bars, while the inner and outer shaded bands indicate systematic uncertainties due to the analysis procedure and total ones (including the hadronic model) respectively.}
\label{fig:He}
\end{figure}

\begin{figure}[htpb]
  \centering
  \centerline{\includegraphics[scale=0.22]{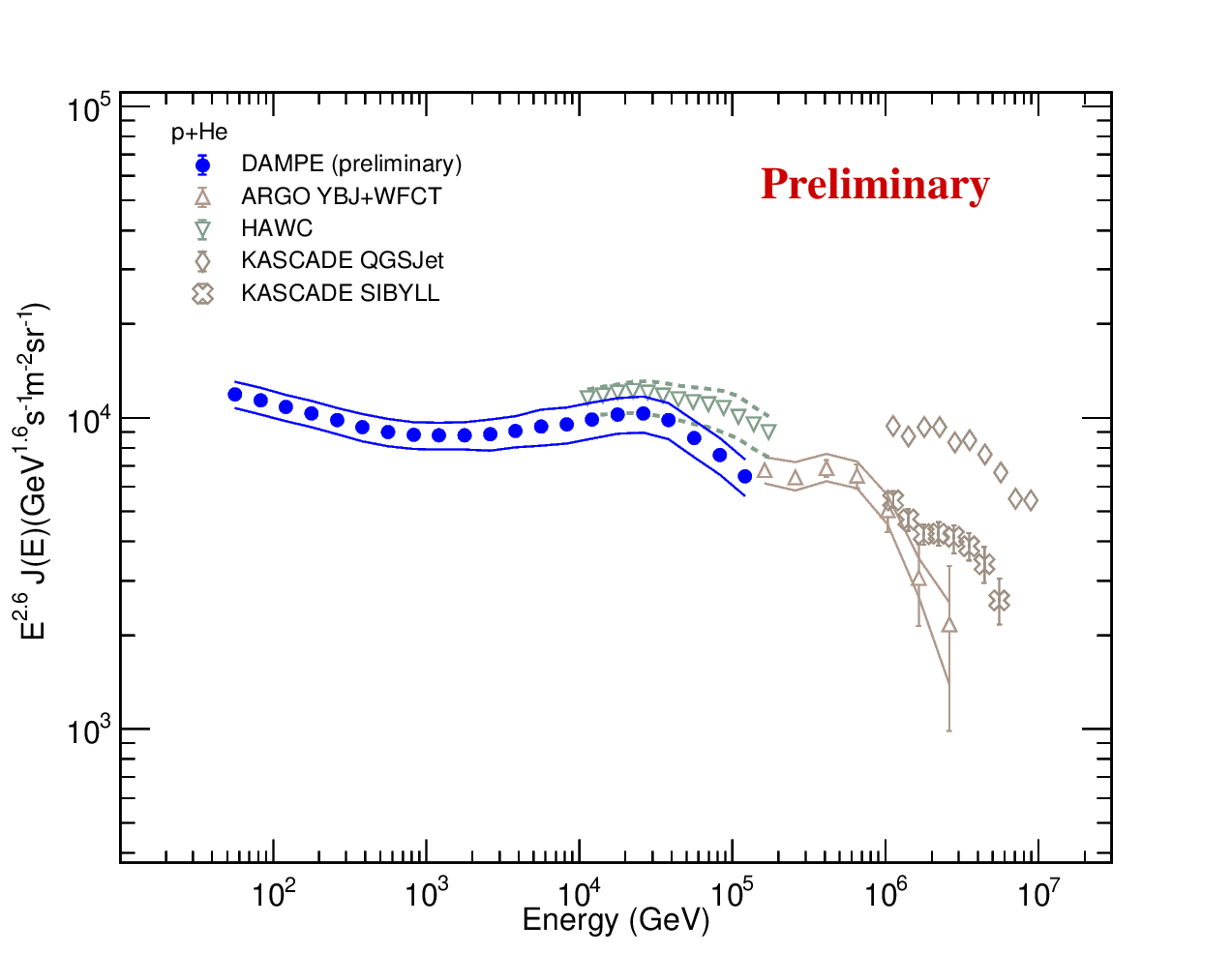}}%
  \caption{Preliminary DAMPE p+He spectrum (blue circles), compared with indirect measurements \protect\cite{18,19,20}, statistical error bars are smaller than the marker size, and the blue band represents the total systematic uncertainties.}
  \label{fig:PrHe}
\end{figure}

\begin{figure}[htpb]
  \centering
  \centerline{\includegraphics[scale=0.26]{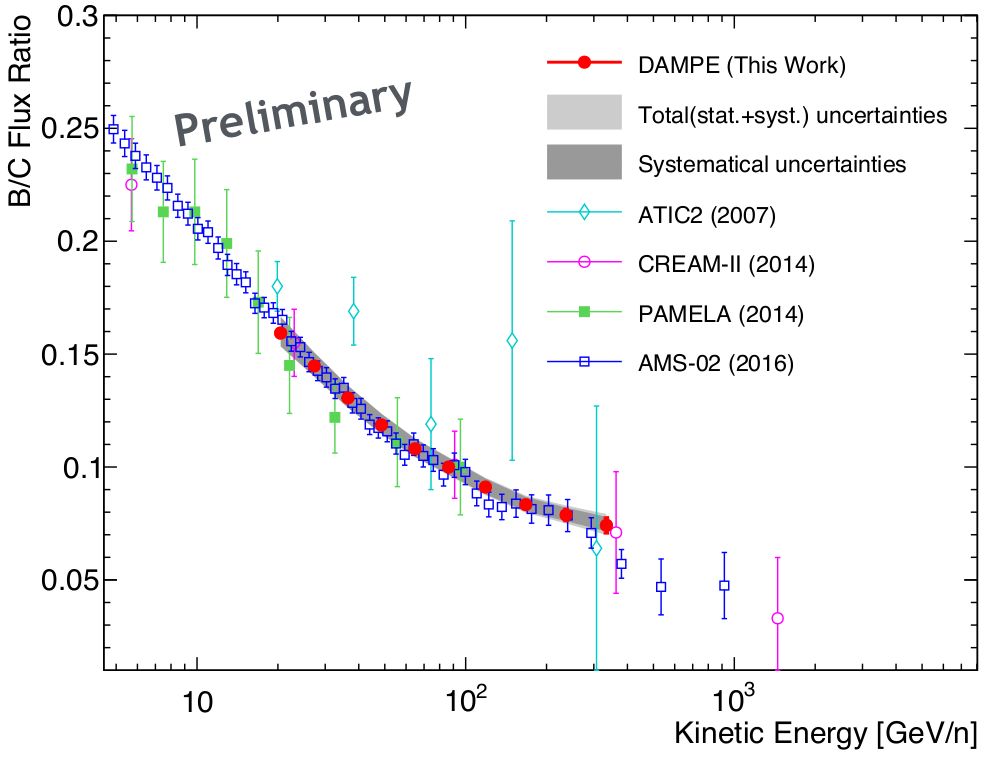}}%
  \caption{Preliminary result on Boron over Carbon flux ratio in red circles, compared with previous results \protect\cite{27,28,29,30}. Statistical uncertainties are represented by error bars, while the grey shaded band represents the quadratic sum of statistical and systematic uncertainties.\label{fig:PROTON_HELIUM}}%
\label{fig:BC}
\end{figure}

Cosmic rays are generated in astrophysical sources, then travel to Earth through the interstellar medium. At 500 km of altitude, DAMPE is able to detect a total of $\sim$5 million CR events per day. In particular, for the cosmic-ray proton flux measurement, 2.5 years of DAMPE data were analyzed in the energy range from 40 GeV to 100 TeV \cite{8}, as illustrated in figure \ref{fig:Pr}. Afterwards, with 4.5 years of DAMPE data, the helium flux with energy from 70 GeV to 80 TeV was measured, as reported in figure \ref{fig:He} \cite{13}. The proton and helium spectra measured by DAMPE confirm the evidence of a spectral hardening, already found by other experiments \cite{1,2,3,4,5,6,7}, and reveal, for the first time, a softening feature at $\sim$14 TeV for protons and at $\sim$34 TeV for helium, with high significance. By selecting a combined proton and helium sample, DAMPE can reach even higher energy and purity, providing a link between direct and indirect cosmic-ray measurements, and giving further confirmation of the hardening and softening features. The preliminary result on the combined proton and helium spectrum can be seen in figure \ref{fig:PrHe} between the energies 50 GeV and 150 TeV \cite{17}. The extension of this spectrum to higher energy is in progress. Ongoing cosmic-ray analyses focus on heavier nuclei such as Li, Be, B, C, N, O, Fe \cite{24,25}, and on secondary to primary ratio (B/C) \cite{26}, shown in figure \ref{fig:BC}. These measurements are intended to clarify the CR propagation mechanism in our galaxy.

\section{SUMMARY}
Since December 2016, the DAMPE satellite has been continuously collecting data with all its subdetectors optimally working. In these years of operation, several results were achieved. The electron-positron spectrum has been measured up to 4.6 TeV, with a precision never achieved before, showing the presence of a spectral break at $\sim$ 1 TeV. A large amount of gamma-ray data were collected, resulting in the identification of 222 gamma-ray sources, mainly AGNs. DAMPE also set a new lower limit for dark matter mass, in the case of mass lower than 100 GeV. Ground-breaking results were obtained in the field of CR, where DAMPE measured the spectra of proton and helium revealing, for the first time, a softening structure at $\sim$ 14 TeV for protons and $\sim$ 34 TeV for helium. When proton and helium are combined in a single spectrum (p+He), the result shows that DAMPE has the potentiality to reach hundreds of TeV, thus building a bridge between space-based and ground-based experiments. The latter analysis focused on reaching higher energies, along with other CR nuclei analysis, are currently in progress. By measuring primary (C, O, Fe, ...) and secondary (Li, Be, B, ...) spectra, DAMPE will provide further insights on CR origin, acceleration and propagation mechanisms in our Galaxy.  

\section*{Acknowledgments}

The DAMPE mission was funded by the strategic priority science and technology projects in space science of Chinese Academy of Sciences. In China the data analysis is supported by the National Key Research and Development Program of China (No. 2016YFA0400200), the National Natural Science Foundation of China (Nos. 11921003, 11622327, 11722328, 11851305, U1738205, U1738206, U1738207, U1738208, U1738127), the strategic priority science and technology projects of Chinese Academy of Sciences (No. XDA15051100), the 100 Talents Program of Chinese Academy of Sciences, the Young Elite Scientists Sponsorship Program by CAST (No. YESS20160196), and the Program for Innovative Talents and Entrepreneur in Jiangsu. In Europe the activities and data analysis are supported by the Swiss National Science Foundation (SNSF), Switzerland, the National Institute for Nuclear Physics (INFN), Italy, and the European Research Council (ERC) under the European Union's Horizon 2020 research and innovation programme (No. 851103).

\end{document}